\documentclass[aps,prl,twocolumn,superscriptaddress,showpacs]{revtex4-1}

\usepackage{graphicx}
\usepackage{dcolumn}
\usepackage{bm}

\begin{document}

\title{Spin-momentum locked polariton transport in the chiral strong coupling regime}

\author{Thibault Chervy}
\affiliation{ISIS \& icFRC, Universit\'e de Strasbourg and CNRS, UMR 7006, F-67000 Strasbourg, France}

\author{Stefano Azzini}
\affiliation{ISIS \& icFRC, Universit\'e de Strasbourg and CNRS, UMR 7006, F-67000 Strasbourg, France}

\author{Etienne Lorchat}
\affiliation{Universit\'e de Strasbourg, CNRS, IPCMS, UMR 7504, F-67000 Strasbourg, France}

\author{Shaojun Wang}
\affiliation{Dutch Institute for Fundamental Energy Research, Eindhoven, The Netherlands}

\author{Yuri Gorodetski}
\affiliation{Mechanical Engineering and Mechatronics Department and Electrical Engineering and Electronics Departement, Ariel University, Ariel 40700, Israel}

\author{James A. Hutchison}
\affiliation{ISIS \& icFRC, Universit\'e de Strasbourg and CNRS, UMR 7006, F-67000 Strasbourg, France}

\author{St\'ephane Berciaud}
\affiliation{Universit\'e de Strasbourg, CNRS, IPCMS, UMR 7504, F-67000 Strasbourg, France}

\author{Thomas W. Ebbesen}
\affiliation{ISIS \& icFRC, Universit\'e de Strasbourg and CNRS, UMR 7006, F-67000 Strasbourg, France}

\author{Cyriaque Genet}
\affiliation{ISIS \& icFRC, Universit\'e de Strasbourg and CNRS, UMR 7006, F-67000 Strasbourg, France}
\email[Corresponding author:]{genet@unistra.fr}

\date{\today}

\begin{abstract}
We demonstrate room temperature chiral strong coupling of valley excitons in a transition
metal dichalcogenide monolayer with spin-momentum locked surface plasmons.
In this regime, we measure spin-selective excitation of directional
flows of polaritons. Operating under strong light-matter coupling, our
platform yields robust intervalley contrasts and coherences, enabling us
to generate coherent superpositions of chiral
polaritons propagating in opposite directions. Our results reveal the rich and easy to implement possibilities offered by our
system in the context of chiral optical networks. 

\end{abstract} 


\maketitle

Optical spin-orbit (OSO) interaction couples the polarization of a
light field with its propagation direction \cite{bliokh2015spin}. An important body of work has recently
described how OSO interactions can be exploited at the level of nano-optical
devices, involving dielectric \cite{bomzon2002space,Kuipers2014,Rauschenbeutel2014science,rafayelyan2016reflective} or plasmonic 
architectures \cite{bliokhPRL2008,gorodetski2013generating,Capasso2013,Zayats2013,spektor2015,Drezet2016}, all able to confine the electromagnetic field
below the optical wavelength. Optical spin-momentum locking effects have been used to
spatially route the flow of surface plasmons depending on
the spin of the polarization of the excitation beam \cite{Zayats2014} or to
spatially route the flow of photoluminescence (PL) depending on
the spin of the polarization of the emitter transition
\cite{Rauschenbeutel2014natcomm}. Such 
directional coupling, also known as chiral coupling, has been 
demonstrated in both the classical and in the quantum regimes \cite{Junge2013,Lodhal2015,GonzalezPRB2015,Rauschenbeutel2015,Young2015,GonzalezPRA2016,Coles2016}. Chiral coupling
opens new opportunities in the field of light-matter interactions
with the design of non-reciprocal devices,
ultrafast optical switches, non destructive photon
detector, and quantum memories and networks (see \cite{Zoller2016} and
references therein).  

In this letter, we propose a new platform consisting of spin-polarized
valleys of a transition metal dichalcogenide (TMD) monolayer strongly 
coupled to a plasmonic OSO mode, at room temperature (RT).
In this strong coupling regime, each spin-polarized valley exciton is
hybridized with a single plasmon mode of specific momentum.
The chiral nature of this interaction generates spin-momentum
locked polaritonic states, which we will refer to with the
portmanteau {\it chiralitons}. 
A striking feature of our platform is its capacity to induce
RT robust valley contrasts, enabling the directional transport of chiralitons over
micron scale distances. Interestingly, the strong coupling regime also yields coherent intervalley dynamics whose contribution can still be observed in the steady-state. We hence demonstrate the generation of coherent superpositions (i.e. pairs) of chiralitons flowing in opposite directions.
These results, unexpected from the bare TMD monolayer RT
properties \cite{Xu2013,moody2016exciton,Li2016}, point towards the importance of the strong coupling regime where fast Rabi oscillations compete with TMD valley relaxation dynamics, as recently discussed \cite{Tartakovskii2016,Chen2017,Kleemann2017}.

\begin{figure}
\includegraphics[width=1\columnwidth]{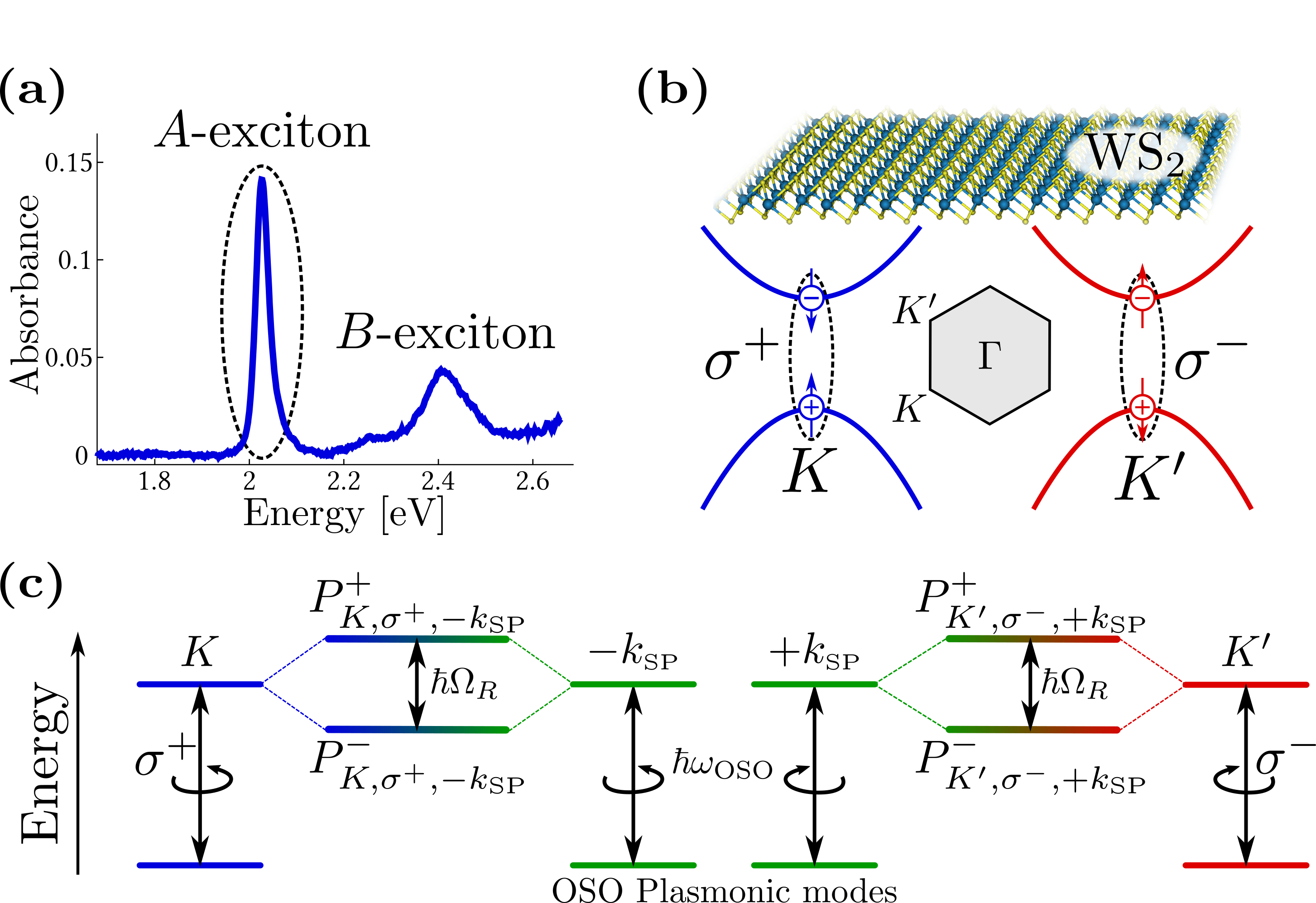}
\caption{(a) Absorbance spectrum of a WS$_2$ monolayer as
  obtained from its transmission spectrum.  
  (b) Crystal packing of a tungsten disulfide (WS$_2$)
  monolayer, and sketch of its electronic
  band structure around the points $K$ and $K'$ of the Brillouin zone, with the
  corresponding optical selection rules for band-edge excitons
  formation under left ($\sigma^+$) and right ($\sigma^-$) circular excitation. 
  (c) Energy level diagram of
  the $K$ and $K'$ excitons of  WS$_2$ strongly coupled to an OSO plasmonic
  mode at energy $\hbar\omega_{\rm OSO}$ and wavevector $\pm k_{\textrm{\tiny SP}}$. 
  \label{fig:1}}
\end{figure}

The small Bohr radii and reduced screening of monolayer TMD excitons
provide the extremely large oscillator strength required for light 
matter interaction in the strong coupling regime, as already achieved
in Fabry-P\'erot cavities \cite{Menon2015,Tartakovskii2015,Imamoglu2016}
and more recently in plasmonic resonators \cite{Agarwal2016,ShaojunWS2}.
In this context, Tungsten Disulfide (WS$_2$) naturally sets itself as
a perfect material for RT strong coupling \cite{ShaojunWS2} due to
its sharp and intense $A$-exciton absorption peak, well separated from
the higher energy $B$-exciton line (see Fig.~\ref{fig:1}(a)) \cite{Heinz2014}. 
Moreover, the inversion symmetry breaking of the crystalline
order on a TMD monolayer,  
combined with time-reversal symmetry, leads to spin-polarized
valley transitions at the $K$ and $K'$ points of the associated
Brillouin zone, as sketched in Fig.~\ref{fig:1}(b) \cite{reviewTMDs}. This polarization property makes therefore atomically thin TMD semiconductors 
very promising candidates with respect to the chiral aspect of the
coupling between the excitonic valleys and the plasmonic OSO modes \cite{LiACS2016,Gong2017},
resulting in the strongly coupled energy diagram shown in Fig.~\ref{fig:1}(c).

\begin{figure}
\includegraphics[width=1\columnwidth]{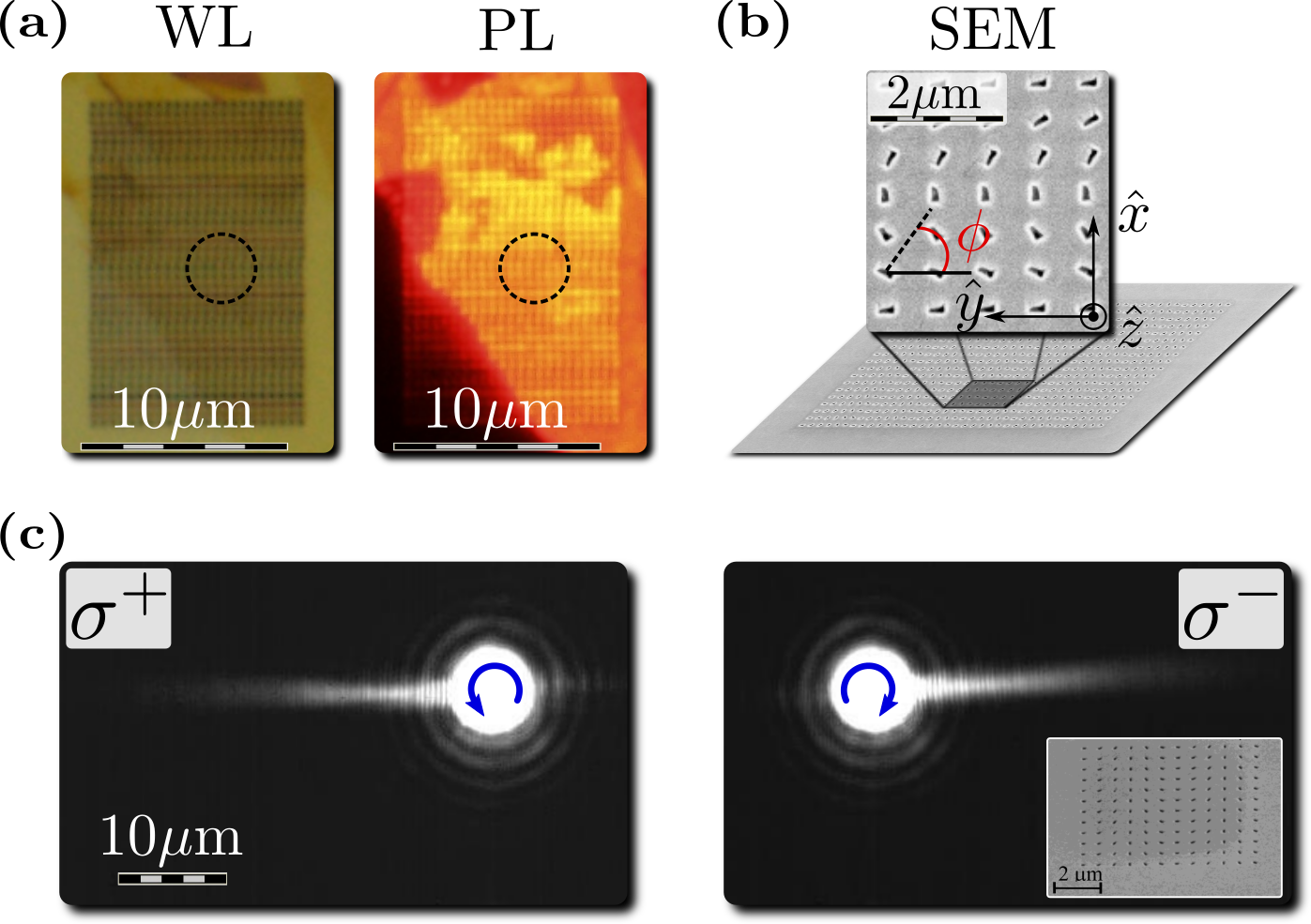}
\caption{(a) White light (WL) microscope image of the sample and
  photoluminescence (PL) image of the same area under $2.58$ eV
  excitation. (b) SEM image of the plasmonic OSO resonator fabricated by sputtering $200$ nm of gold on a glass substrate coated by a
$5$ nm thick chromium adhesion layer. The array with $\phi$-rotated rectangular apertures is milled
through the metallic layers using a focused ion beam
(FIB). (c) Real-space leakage radiation microscope \cite{Drezet2016} images of the 
  surface plasmons launched by $\sigma^+$ and $\sigma^-$ excitations
  on a OSO plasmonic resonator similar to the one of panel (b). \label{fig:2}}    
\end{figure}

Experimentally, our system, shown in
Fig.~\ref{fig:2}(a), consists of a mechanically exfoliated
monolayer of WS$_2$ covering a
plasmonic OSO hole array, with a $5$ nm thick dielectric spacer (polymethyl
methacrylate). 
The array, imaged in Fig.~\ref{fig:2}(b), is designed on a $(x,y)$
square lattice with a grating period $\Lambda$, and consists of rectangular
nano-apertures $(160\times90$ nm$^2)$ rotated stepwise along the
$x$-axis by an angle $\phi=\pi/6$. The associated orbital period
$6\times\Lambda$ sets a rotation vector $\boldsymbol{\Omega}=(\phi /
\Lambda)\hat{z}$, which combines with the spin $\sigma$ of the
incident light into a geometric phase $\Phi_g=- \Omega\sigma x$
\cite{bliokh2008coriolis}. The gradient of this geometric phase
imparts a  momentum ${\bf k}_g=-\sigma(\phi / \Lambda)\hat{x}$
added to the matching condition on the array between the
plasmonic ${\bf k}_{\rm SP}$ and incidence in-plane ${\bf
  k}_{\rm in}$ momenta: 
${\bf k}_{\rm
  SP}={\bf k}_{\rm in} + (2\pi / \Lambda) (n\hat{x}+m\hat{y})+{\bf
  k}_g$.
This condition defines different $(n,m)$ orders for the plasmonic
dispersions, which are transverse magnetic (TM) and transverse electric (TE)
polarized along the $x$ and $y$ axis 
of the array respectively (see Fig.~\ref{fig:2}(b)).
The dispersive properties of such a resonator thus combines two modal responses: plasmon excitations directly determined on the square Bravais lattice of the grating for both
$\sigma^+$ and $\sigma^-$ illuminations via $(2\pi / \Lambda) (n\hat{x}+m\hat{y})$, and spin-dependent plasmon OSO modes launched by the additional geometric momentum ${\bf k}_{g}$.
It is important to
note that the contribution of the geometric phase impacts the TM
dispersions only. The period of our structure $\Lambda=480$ nm is
optimized to have  $n=+1$ and $n=-1$ TM modes resonant with the  
absorption energy of the $A$-exciton of WS$_2$ at $2.01$ eV for
$\sigma^+$ and $\sigma^-$ illuminations respectively. This
strict relation between $n=\pm
1$ and $\sigma=\pm 1$ is the OSO mechanism that breaks the left 
vs. right symmetry of the modal response of the array, which in this
sense becomes chiral.  Plasmon OSO modes are thus launched in
counter-propagating directions along the $x$-axis for
opposite spins $\sigma$ of 
the excitation light. In the case of a bare plasmonic OSO resonator,
this is clearly observed in Fig.~\ref{fig:2} (c). 
We stress that similar arrangements of anisotropic apertures have previously been
demonstrated to allow for spin-dependent surface plasmon
launching \cite{Hasman2011,SZhang2013,Capasso2013,Drezet2016}. 

\begin{figure}
\includegraphics[width=1\columnwidth]{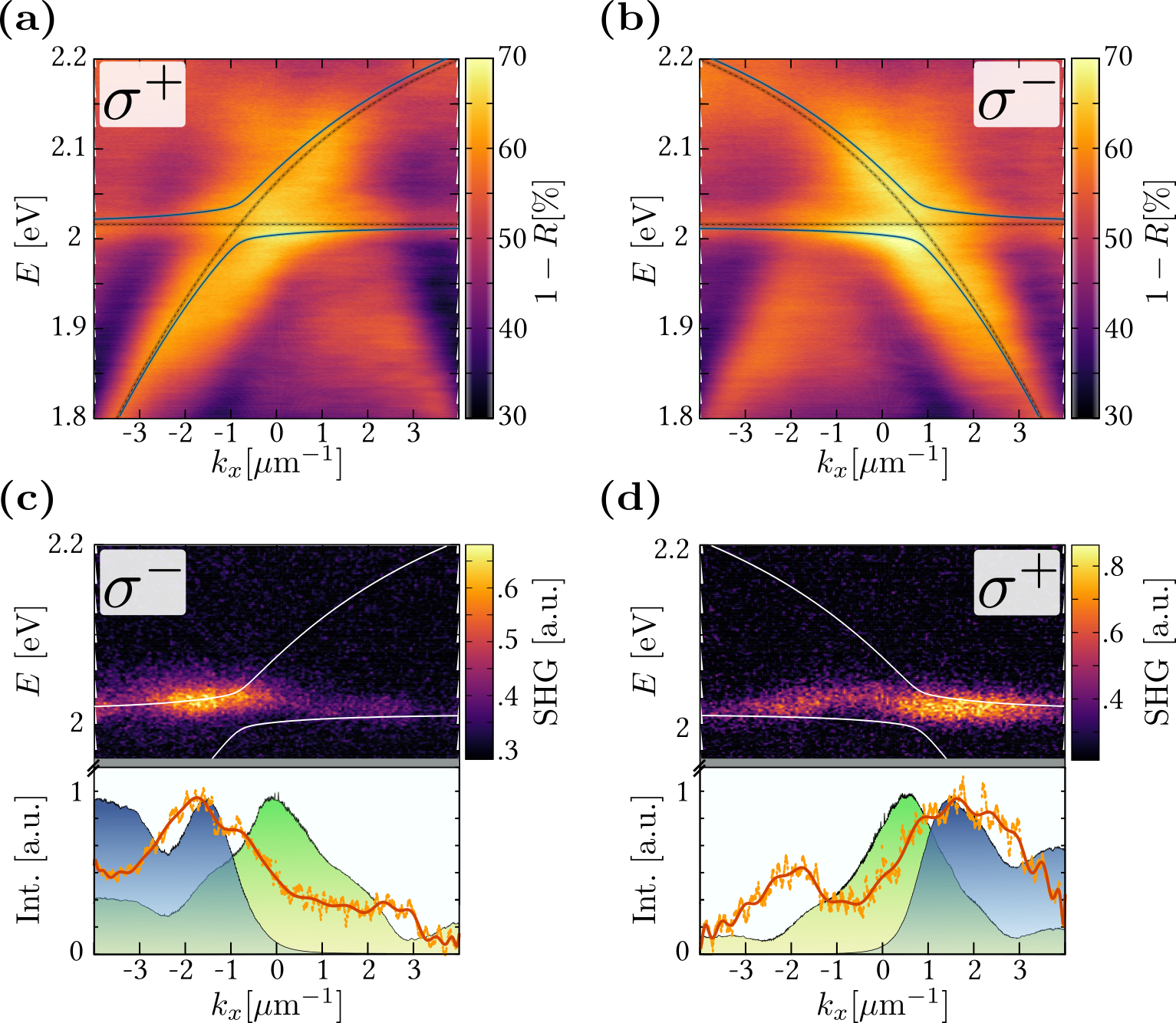}
\caption{Angle-resolved absorption spectrum of the sample analyzed in
  (a) left and (b) right circular polarizations, with the best fit coupled oscillator model drawn. Angle-resolved resonant second harmonic spectrum  for (c) right and (d) left circular excitations at $1.01$eV. Corresponding crosscuts are displayed with the angular profile of the SH signal (red curves), of the absorption spectra at $2$ eV (green shades) and of the product (blue shades) between the absorption and the $4^{\rm th}$ power of the excitonic Hopfield coefficient of the chiralitonic state -see details in the Supporting Information (Sec. D)
   \label{fig:3}.}
\end{figure}

As explained in the Supporting Information (Sec. A), the low transmission measured through our WS$_2$/plasmonic array
sample (Fig.~\ref{fig:2}(a)) enables us to obtain absorption spectra directly
from reflectivity spectra.
Angle-resolved white light absorption spectra are hence recorded and shown in Fig.~\ref{fig:3} (a) and (b) for left and
right circular polarizations. In each case, two strongly dispersing branches are
observed, corresponding to upper and lower chiralitonic
states. As detailed in the Supporing Information (Sec. A), a fit of a coupled dissipative oscillator model to the dispersions enables us to extract a branch splitting $2\sqrt{\left(\hbar\Omega_R\right)^2 - \left( \hbar\gamma_{\tiny\textrm{ex}}- \hbar\Gamma_{\tiny\textrm{OSO}}\right)^2}=40$ meV. With measured linewidths of the excitonic mode $\hbar\Gamma_{\tiny\textrm{OSO}}=80$ meV and of the plasmonic mode $\hbar\gamma_{\tiny\textrm{ex}}=26$ meV, this fitting yields a Rabi frequency 
of $\hbar\Omega_{\rm R}=70$ meV, close to our previous
observations on non-OSO plasmonic resonators \cite{ShaojunWS2}. We emphasize that this value clearly fulfills the strong coupling criterion with a figure-of-merit $\Omega_{\rm R}^2/(\gamma_{\rm exc}^2+\Gamma_{\rm OSO}^2) =0.69$ larger than the $0.5$ threshold that must be reached for strong coupling \cite{Houdre2005,TormaBarnes2015}. This demonstrates that our system does operate in the strong coupling regime, despite the relatively low level of visibility of the anti-crossing. This is only due $(i)$ to spatial and spectral disorders which leave, as always for collective systems,  an inhomogeneous band of uncoupled states at the excitonic energy, and $(ii)$ to the fact that an uncoupled Bravais plasmonic branch is always superimposed to the plasmonic OSO mode, leading to asymmetric lineshapes clearly seen in Fig.~\ref{fig:3} (a) and (b). As shown in the Supporting Information (Sec. A), the anti-crossing can actually be fully resolved through a first-derivative analysis of our absorption spectra.

In such strong coupling conditions,
the two dispersion diagrams also show a clear mirror symmetry breaking with respect to
the normal incident axis ($k_x=0$) for the two opposite optical spins. This clearly
demonstrates the capability of our structure to act as a spin-momentum locked polariton
launcher. From
the extracted linewidth that gives the lifetime of the chiralitonic mode and the curvature of the dispersion relation that provides its group velocity, we can
estimate a chiraliton propagation length of the order of $4~\mu$m, in
good agreement with the measured PL diffusion length presented in the Supporting Information (Sec. B).

In view of chiral light-chiral matter interactions, we further
investigate the interplay between this 
plasmonic chirality and the valley-contrasting chirality of the WS$_2$
monolayer. A first demonstration of such an interplay is found in the
resonant second harmonic (SH) response of the strongly coupled system. 
Indeed, monolayer TMDs have been shown to give a high valley contrast in the
generation of a SH signal resonant with their $A$-excitons
\cite{Seyler2015}. As we show in Supporting Information (Sec. G) such high SH valley contrast are measured on a bare WS$_2$ monolayer. The optical selection rules for SH generation are opposite 
to those in linear absorption since the process involves two
excitation photons, and are more robust since the SH process
is instantaneous. 

The angle resolved resonant SH signals are shown in
Fig.~\ref{fig:3} (c) and (d) for right and left circularly polarized excitation. The SH signals are angularly exchanged when the spin of the excitation is reversed with a right vs. left contrast (ca. $20\%$) close to the one measured on
the reflectivity maps (ca. $15\%$). This unambigously demonstrates the selective coupling of excitons in
one valley to surface plasmons propagating in one direction, thus
realizing valley-contrasting chiralitonic states with spins locked to
their propagation wavevectors:
\begin{eqnarray}  
  &&|P^\pm_{K,\sigma^+,-k_{\textrm{\tiny SP}}}> = |g_K,1_{\sigma^+},-k_{\textrm{\tiny SP}}> \pm |e_K,0_{\sigma^+},-k_{\textrm{\tiny SP}}>   \nonumber\\  
  &&|P^\pm_{K',\sigma^-,+k_{\textrm{\tiny SP}}}> = |g_{K'},1_{\sigma^-},+k_{\textrm{\tiny SP}}> \pm |e_{K'},0_{\sigma^-},+k_{\textrm{\tiny SP}}>,   \nonumber  
\end{eqnarray}
where $e_{i}(g_{i})$ corresponds to the presence (absence) of an
exciton in the valley $i=(K,K')$ of WS$_2$, and $1_{j}(0_{j})$ to 1 $(0)$ plasmon
in the mode of polarization $j=(\sigma^+,\sigma^-)$ and wavevector 
$\pm k_{\textrm{\tiny SP}}$.

The detailed features of SH signal (crosscuts in Fig.~\ref{fig:3} (c) and (c)) reveal within the bandwidth of our pumping laser the contributions of both the uncoupled excitons and the upper chiraliton to the SH enhancement. The key observation, discussed in the Supporting Information (Sec. D), is the angular dependence of the main SH contribution. This contribution, shifted from the anticrossing region, is a feature that gives an additional proof of the strongly coupled nature of our system because it is determined by the excitonic Hopfield coefficient of the spin-locked chiralitonic state. In contrast, the residual SH signal related to the uncoupled (or weakly coupled) excitons simply follows the angular profile of the absorption spectra taken at $2$ eV, thus observed over the anticrossing region. Resonant SH spectroscopy of our system therefore confirms the presence of the chiralitonic states, with the valley contrast
of WS$_2$ and the spin-locking property of the OSO plasmonic resonator
being imprinted on these new eigenstates of the system.

\begin{figure}
\includegraphics[width=1\columnwidth]{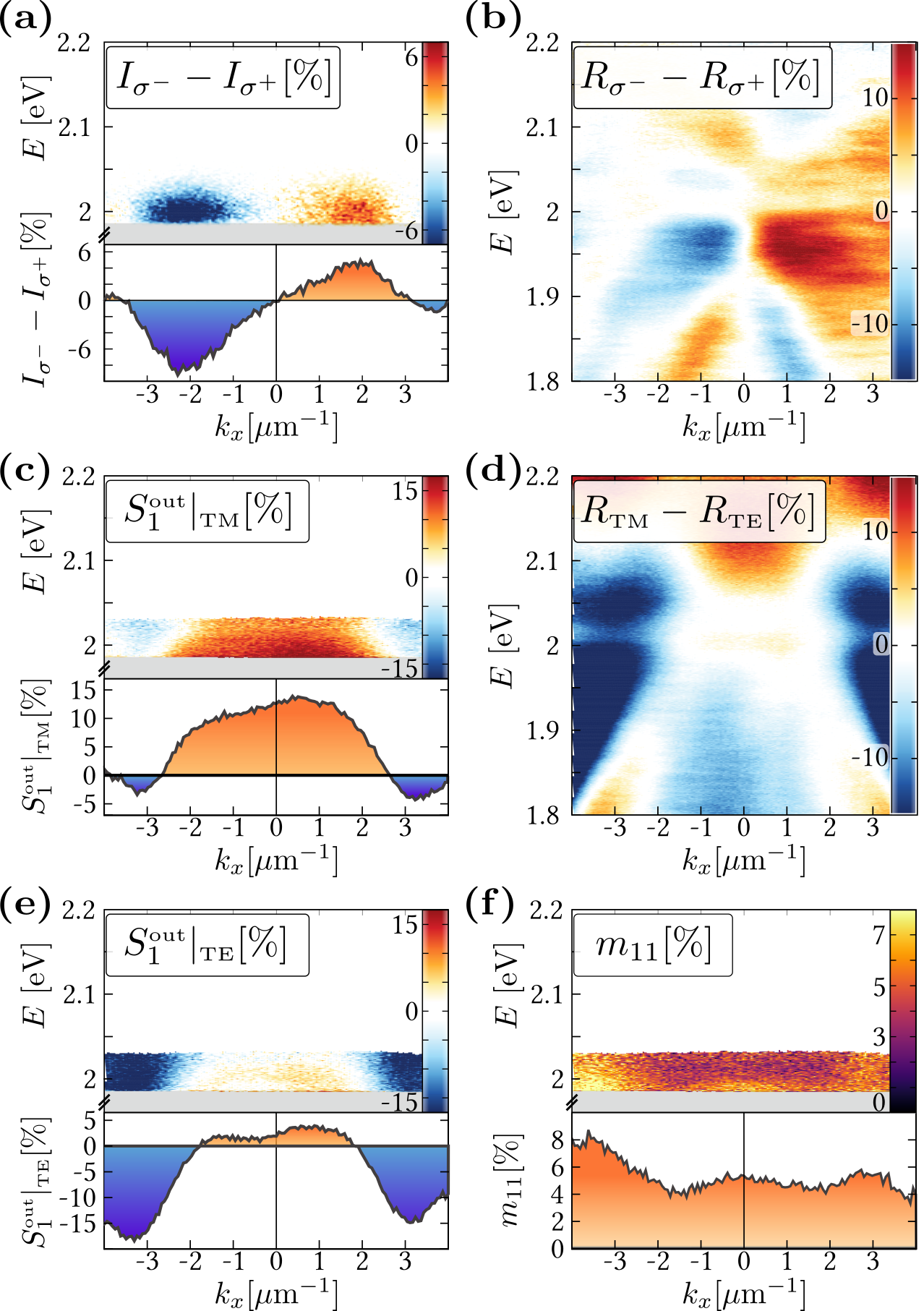}
\caption{(a) Differential PL dispersion spectrum for left and right circularly polarized
  excitations. The shaded regions in all
  panels are removed by the laser line filter, and the cross-cuts are
  taken at $2$ eV. (b) Differential 
  angle-resolved reflection spectrum for left and right circularly
  polarized light. (c), (e) Angle-resolved spectrum of the normalized coefficient $S_1^{\textrm{\tiny out}}|_{\textrm{\tiny TM(TE)}}/S_0$ of
  the PL Stokes vector for a TM(TE) polarized excitation (see text
  for details). Note that we have put a detection threshold below $100$
  photon counts that cuts the signal above $\sim 2.03$ eV in panels
  (a), (c), (e) and (f). (d) Differential
  angle-resolved reflection spectrum obtained from analyzed TM
  and TE measurements. (f) $k_x$-energy dispersion of the degree of
  chiralitonic inter-valley coherence $m_{11}$ computed from (c) and (e). \label{fig:4}}  
\end{figure}

Revealed by these resonant SH measurements, the spin-locking property of chiralitonic states incurs however 
different relaxation mechanisms through the dynamical
evolution of the chiralitons. In particular, excitonic intervalley
scattering can erase valley contrast in WS$_2$ at RT
\cite{Jonker2016} -see below. In our configuration, this would transfer
chiraliton population 
from one valley to the other, generating via optical spin-locking, a
reverse flow, racemizing the chiraliton
population. This picture however does not 
account for the possibility of more robust valley contrasts in
strong coupling conditions, as recently reported with MoSe$_2$ in
Fabry-P\'erot cavities \cite{Tartakovskii2016}.  
The chiralitonic flow is measured by performing angle resolved polarized PL
experiments, averaging the signal over the PL lifetime of ca. $200$
ps (see Supporting Information, Sec. D and E). For these
experiments, the laser excitation
energy is chosen at $1.96$ eV, slightly below the WS$_2$ band-gap. At this energy, the measured PL results
from a phonon-induced up-convertion process that minimizes intervalley scattering events \cite{Xiaodong2016}.
The difference between PL dispersions obtained with left and right
circularly polarized excitations is displayed in Fig.~\ref{fig:4} (a),
showing net flows of chiralitons with 
spin-determined momenta. This is in agreement with the
differential white-light reflectivity map 
$R_{\sigma^-}-R_{\sigma^+}$ of Fig.~\ref{fig:4} (b). Considering that this map gives the sorting efficiency of our OSO resonator, such correlations in the PL implies that the effect of the initial spin-momentum determination of the chiralitons (see Fig.~\ref{fig:3} (e) and (f)) is still observed after $200$ ps at RT. 
After this PL lifetime, a net chiral flow
$\mathcal{F}=I_{\sigma^-}-I_{\sigma^+}$ of  $\sim 6\%$ is extracted
from Fig.~\ref{fig:4} (a). This is the signature of a chiralitonic
valley polarization, in striking contrast with the absence of 
valley polarization that we report for a bare WS$_2$ monolayer at RT
in the Supporting Information, Sec. G.
The extracted net flow is however limited by the 
finite optical contrast $\mathcal{C}$ of our OSO resonator, which we
measure at a $15\%$ level from a cross-cut taken on 
Fig.~\ref{fig:4} (b) at $1.98$ eV. It is therefore possible to
  infer that a chiralitonic valley contrast of 
$\mathcal{F} / \mathcal{C} \simeq 40\%$ can be reached at RT for the strongly coupled WS$_2$ monolayer. 
As mentioned above, we understand this surprisingly robust contrast by invoking the fact that under strong coupling conditions, valley relaxation is outweighted by the faster Rabi energy exchange between the exciton of each valley and the corresponding plasmonic OSO mode, as described in the Supporting Information (Sec. A). From the polaritonic point of view, the local dephasing and scattering processes at play on bare excitons -that erase valley contrasts on a bare WS2 flake as observed in the Supporting Information (Sec. G)- are reduced by the delocalized nature of the chiralitonic state, a process akin to motional narrowing and recently observed on other polaritonic systems \cite{Whittaker1996, Tartakovskii2016,Chen2017,Kleemann2017}.

As a consequence of this motional narrowing effect, such a strongly coupled system involving atomically thin crystals of
TMDs could then provide new ways to incorporate intervalley coherent
dynamics \cite{Xu2013,Li2016,Urbaszek2016,schmidt2016magnetic,ye2017optical} into the 
realm of polariton physics. To illustrate this, we now show
that two counter-propagating flows of chiralitons can evolve coherently. It is clear from Fig.~\ref{fig:1} (c) that within such a coherent superposition of counter-propagating chiralitons
\begin{equation}
  |\Psi> = |P^{\pm}_{K,\sigma^+,-k_{\textrm{\tiny SP}}}> + |P^{\pm}_{K',\sigma^-,+k_{\textrm{\tiny SP}}}>
 \label{eq:Psi}
\end{equation} 
flow directions and spin polarizations become non-separable.
Intervalley coherence is expected to result in a non-zero
degree of linearly polarized 
PL when excited by the same linear polarization. This can be monitored
by measuring the $S_1=I_{\textrm{\tiny TM}}-I_{\textrm{\tiny TE}}$
coefficient of the PL Stokes vector, where 
$I_{\textrm{\tiny TM}(\textrm{\tiny TE})}$ is the emitted PL intensity
analyzed in TM (TE) polarization. 
This coefficient is displayed in the $k_x$-energy plane
in Fig.~\ref{fig:4}(c) for an incident
TM polarized excitation at $1.96$ eV. Fig.~\ref{fig:4}(e) displays the
same coefficient under TE excitation. 
A clear polarization anisotropy on the
chiraliton emission is observed for both TM and TE excitation
polarizations, both featuring the same symmetry along the $k_x=0$ axis
as the differential reflectivity dispersion map 
$R_{\textrm{\tiny TM}}-R_{\textrm{\tiny TE}}$ shown in
Fig.~\ref{fig:4}(d). As detailed in the
Supporting Information (Sec. F), the degree
of chiralitonic intervalley coherence can be directly quantified by the
difference $(S_1^{\textrm{\tiny out}}|_{\textrm{\tiny
    TM}}-S_1^{\textrm{\tiny out}}|_{\textrm{\tiny TE}})/2$, which
measures the PL linear depolarization factor displayed (as $m_{11}$) in
Fig.~\ref{fig:4} (f). By this procedure, we retrieve a chiralitonic
intervalley coherence that varies between $5\%$ and $8\%$ depending on
$k_x$. Interestingly, these values that we reach at RT have magnitudes comparable to those reported on a bare WS$_2$ monolayer at 10 K \cite{Xiaodong2014}. This
unambiguously shows how such strongly coupled TMD systems can sustain
RT coherent dynamics robust enough to be observed despite the long exciton PL lifetimes and plasmonic propagation distances. 

In summary, we demonstrate valley contrasting 
spin-momentum locked chiralitonic states in an atomically thin TMD
semiconductor strongly coupled to a plasmonic OSO resonator. Likely, the observation of such contrasts even after 200 ps lifetimes is made possible by the unexpectedly robust RT coherences inherent to the strong coupling regime. Exploiting such robust coherences, we measure chiralitonic
flows that can evolve in superpositions over micron scale
distances.
Our results show that the combination of
OSO interactions with TMD valleytronics is an interesting path to
follow in order to explore and manipulate RT coherences in chiral quantum architectures \cite{Coles2016,low2016polaritons}.

\begin{acknowledgments}
We thank David Hagenm\"uller for fruitful discussions.
 This work was supported in part by the ANR Equipex ``Union'' (ANR-10-EQPX-52-01), ANR Grant (H2DH
 ANR-15-CE24-0016), the Labex NIE projects
 (ANR-11-LABX-0058-NIE) and USIAS within the Investissement d'Avenir
 program ANR-10-IDEX-0002-02. Y. G. acknowledges support from the
 Ministry of Science, Technology and Space, Israel. S. B. is a member
 of the Institut Universitaire de France (IUF).
\end{acknowledgments}

{\it Author Contributions -}
T. C. and S. A. contributed equally to this work.

\bibliography{biblio-new-arxiv-v2}

\newpage

\section{Supporting Information}

\section{A: Linear absorption dispersion analysis}

Angle resolved absorption spectra  
are given from the measured reflectivity of
the WS$_2$ flake on top of the plasmonic grating
$R_{\tiny\textrm{sample}}$ with:
\begin{equation}
  A = 1 - R_{\tiny\textrm{sample}}/R_{\tiny\textrm{substrate}},
\end{equation}
where $R_{\tiny\textrm{substrate}}$ is the angle resolved reflectivity of
the optically thick (200 nm thickness) Au substrate. 
The $1\%$ max. transmission through the structure
 can be safely neglected.
As we explain in the main text, the resulting dispersion spectra
are broadened by the contribution of different plasmonic modes as well
as coupled and uncoupled exciton populations.

\begin{figure}
\includegraphics[width=1\columnwidth]{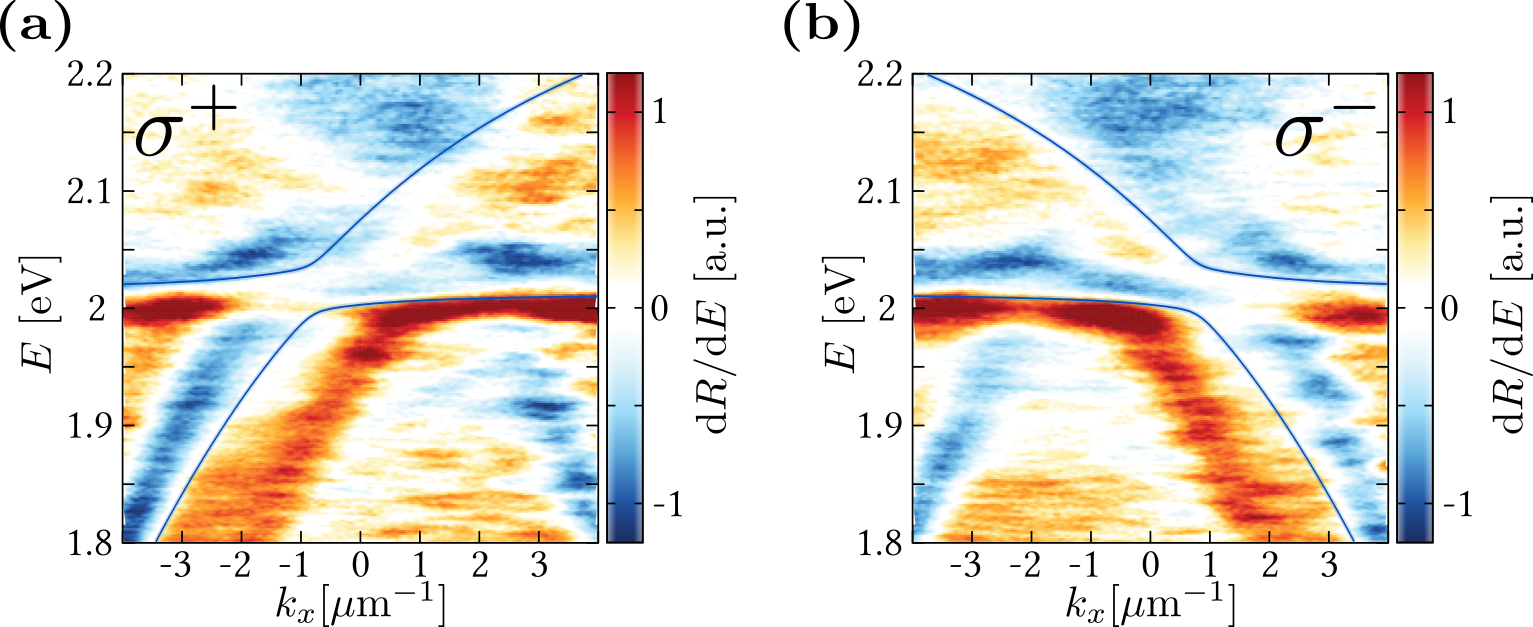}
\caption{First derivative reflectivity maps in (a) left and (b) right circular polarizations, with the best fit coupled oscillator model drawn.   \label{fig:S1}} 
\end{figure}

In order to highlight the polaritonic contribution in the absorption spectra, we
calculate the first derivative of the reflectivity
dispersions $\textrm{d} \left[
R_{\tiny\textrm{sample}}/R_{\tiny\textrm{substrate}} \right] / \textrm{d}E$.  The
derivative was approximated by interpolating the reflectivity spectra on
an equally spaced wavelength grid of step $\Delta \lambda = 0.55$ nm
and using the following finite difference expression
valid up to fourth order in the grid step:
\begin{equation}
  \frac{\textrm{d}R}{\textrm{d}\lambda}(\lambda_0) \simeq \frac{\frac{1}{12}R(\lambda_{-2})
    -\frac{2}{3}R(\lambda_{-1}) + \frac{2}{3}R(\lambda_{+1}) -
    \frac{1}{12}R(\lambda_{+2})}{\Delta \lambda}, 
\end{equation}
where $R(\lambda_{n})$ is the reflectivity evaluated $n$ steps away
from $\lambda_{0}$. The resulting first derivative reflectivity spectra were then converted to an
energy scale and plotted as dispersion diagrams in Fig.~\ref{fig:S1} (a) and (b).

In these first derivative reflectivity maps, the zero-crossing points correspond
to the peak positions of the modes, and the maxima and minima
indicate the inflection points of the reflectivity lineshapes. 
At the excitonic asymptotes of the dispersion curves, where the
polaritonic linewidth is expected to match that of the bare WS$_2$
exciton, we read a linewidth of $26$ meV from the maximum to
minimum energy difference of the derivative reflectivity maps.
This value is equal to the full width at half maximum (FWHM) $\hbar\gamma_{\rm exc}$ that we measured
from the absorption spectrum of a bare WS$_2$ flake on a dielectric
substrate.

On the low energy plasmonic asymptotes, we clearly observe the effect
of the Bravais and OSO modes, partially overlapping in an asymmetric
broadening of the branches. In this situation, a measure of the mode half-widths can
be extracted from the (full) widths of the positive or negative regions of the first
differential reflectivity maps. This procedure yields an energy
half-width for the plasmonic modes of $\hbar\Gamma_{\rm OSO}/2=40$ meV. This width in energy
can be related to an in-plane momentum width of ca. $0.5
~\mu\textrm{m}^{-1}$ via the plasmonic group velocity $v_G 
= \partial E/\partial k = 87$
meV$\cdot\mu$m, that we calculate from the branch curvature at $1.85$ eV.
This in-plane momentum width results in a plasmonic propagation length
of about $4 ~\mu$m. This value is in very good agreement with the measured PL
extension above the structure, as discussed in section B below, validating our estimation of the mode linewidth $\hbar\Gamma_{\rm OSO}=80$ meV. 

The dispersive modes of the system can be modeled by a dipolar
Hamiltonian, where excitons in each valley are selectively 
coupled to degenerated OSO plasmonic modes of opposite wavevectors
$\pm {\bf k}_{\rm SP}$, as 
depicted in Fig.~2 in the main text: 
\begin{equation}
  \mathcal{H} = \sum_{k_x}\left[\mathcal{H}_{\tiny\textrm{OSO}}(k_x) +
  \mathcal{H}_{\tiny\textrm{ex}} +
  \mathcal{H}_{\tiny\textrm{int}}(k_x)\right], \label{eq:H}
\end{equation}
which consists of three different contributions:
\begin{equation}
  \mathcal{H}_{\tiny\textrm{OSO}}(k_x) =
  \hbar\omega_{\tiny\textrm{OSO}}(k_x)\left(a^\dagger_{k_x}a_{k_x} +
    a^\dagger_{-k_x}a_{-k_x}\right), 
\end{equation}
\begin{equation}
  \mathcal{H}_{\tiny\textrm{ex}}(k_x) =
  \hbar\omega_{\tiny\textrm{ex}}\left(b^\dagger_{K-k_x}b_{K-k_x} + b^\dagger_{K'+k_x}b_{K'+k_x}\right),
\end{equation}
\begin{eqnarray}
  \mathcal{H}_{\tiny\textrm{int}}(k_x) &=& \hbar g\left(a^\dagger_{k_x}
      + a_{k_x}\right)\left(b^\dagger_{K'+k_x} + b_{K'+k_x}\right) \nonumber \\
   & & + \hbar g\left(a^\dagger_{-k_x}
      + a_{-k_x}\right)\left(b^\dagger_{K-k_x} + b_{K-k_x}\right),
\end{eqnarray}
where $a(a^\dagger)$ are the lowering (raising) operators of the
OSO plasmonic modes of energy $\hbar\omega_{\tiny\textrm{OSO}}(k_x)$,
$b(b^\dagger)$ are the lowering (raising) operators of the 
exciton fields of energy $\hbar\omega_{\tiny\textrm{ex}}$, and $g = \Omega_R/2$ is the
light-matter coupling frequency. In this hamiltonian the chiral
light-chiral matter interaction is effectively accounted for by
coupling excitons of the valley $K'(K)$ to plasmons propagating with
wavevectors $k_x(-k_x)$. Moreover, the dispersion of
the exciton energy can be neglected on the scale of the plasmonic wavevector ${\bf k}_{\rm SP}$.

Using the Hopfield procedure \cite{CiutiPRB2005}, we can diagonalize
the total Hamiltonian by 
finding polaritonic normal mode operators $P^\pm_{K(K')}$ associated with each
valley, and obeying the following equation of motion at each $k_x$
\begin{equation}
  \left[P^\pm_{K(K')},\mathcal{H}\right] = \hbar\omega_\pm
  P^\pm_{K(K')}, \label{eq:motion}
\end{equation}
with $\omega_\pm>0$. In the rotating wave approximation (RWA), justified
here by the moderate coupling strength (see below), $P^j_{\lambda}
\simeq \alpha^j_{\lambda}a +\beta^j_{\lambda}b$, $j\in\{+,-\}$ and $\lambda\in\{K,K'\}$.
The plasmonic and excitonic Hopfield coefficients $\alpha^j_{\lambda}(k_x)$
and $\beta^j_{\lambda}(k_x)$ are obtained by diagonalizing the following
matrix for every $k_x$
\begin{equation}
\left(
\begin{array}{cccc}
\hbar\omega_{\tiny\textrm{OSO}} & i\hbar\Omega_{\rm R} & 0 & 0 \\ 
-i\hbar\Omega_{\rm R} & \hbar\omega_{\tiny\textrm{ex}} & 0 & 0 \\ 
0 & 0 & -\hbar\omega_{\tiny\textrm{OSO}} & i\hbar\Omega_{\rm R}\\ 
0 & 0 & -i\hbar\Omega_{\rm R}& -\hbar\omega_{\tiny\textrm{ex}}
\end{array} 
\right). \label{eq:Matrix}
\end{equation}

The dynamics of the coupled system will be ruled by the
competition between the coherent evolution described by the Hamiltonian (\ref{eq:H}) and the
different dissipative processes contributing to the uncoupled modes
linewidths. This can be taken into account by including the measured
linewidths as imaginary parts in the excitonic and plasmonic mode
energies (Weisskopf-Wigner approach). Under such conditions, we evaluate the eigenvalues $\omega_{\pm}$ of the the matrix (\ref{eq:Matrix}). The real parts of $\omega_{\pm}$ are then fitted to the maxima of the angle resolved reflectivity maps presented on Fig.~3 in the main text, or to the zeros of the first derivative reflectivity maps
shown here in Fig.~\ref{fig:S1} (a) and (b). Both procedures give the same best fit that yields the polaritonic branch splitting as \cite{Houdre2005,TormaBarnes2015}
\begin{equation}
\hbar(\omega_+-\omega_-) = 2\sqrt{\left(\hbar\Omega_R\right)^2 - \left( \hbar\gamma_{\tiny\textrm{ex}}- \hbar\Gamma_{\tiny\textrm{OSO}}\right)^2}
\label{eq:SCsplit}
\end{equation}
which equals $40$ meV. From the determination (see above) of the FWHM of the excitonic $\hbar\gamma_{\tiny\textrm{ex}}$ and plasmonic $\hbar\Gamma_{\tiny\textrm{OSO}}$ modes, we evaluate a Rabi energy $ \hbar\Omega_R=70$ meV. 

These values give a ratio
\begin{equation}
  \left(\hbar\Omega_R\right)^2 / \left(\left(\hbar\gamma_{\tiny\textrm{ex}}\right)^2 +
 \left(\hbar  \Gamma_{\tiny\textrm{OSO}}\right)^2\right) = 0.69,
\end{equation}
above the $0.5$ threshold which is the strong coupling criterion -see \cite{Houdre2005} for a detailed discussion. 
This figure-of-merit of $0.69>0.5$ therefore clearly demonstrates that our system is operating in the strong coupling regime. 

Interestingly, the intervalley scattering rate $\hbar\gamma_{KK'}$ does not
enter in this strong coupling criterion. Indeed, such
events corresponding to an inversion of the valley indices
$K\leftrightarrow K'$ do not contribute to 
the measured excitonic linewidth, and are thus not detrimental to the
observation of strong coupling. In the $\hbar\Omega_R \ll
\hbar\gamma_{KK'}$ limit, the Hamiltonian (\ref{eq:H}) would reduce to the usual RWA
Hamiltonian and the valley contrasting chiralitonic behavior would be lost.  
The results gathered in Fig.~4 in the main text clearly show that this is not the case for our system, 
allowing us to conclude that the Rabi frequency overcomes such
intervalley relaxation rates. Remarkably, strong coupling thus
allows us to put an upper bounds to those rates, in close relation with \cite{Holmes}. 
 
\section{B: Chiraliton diffusion length}

The diffusion length of chiralitons can be estimated by measuring the
extent of their photoluminescence (PL) under a tightly focused
excitation. To measure this extent, we excite a part of a WS$_2$
monolayer located above the plasmonic hole array (Fig. S\ref{fig:S1}(a)
and (b)). This measurement is done on a home-built PL microscope,
using a $100\times$ microscope  
objective of 0.9 numerical aperture and exciting the PL with a HeNe laser at
$1.96$ eV, slightly below the exciton band-gap. A diffraction-limited
spot of $430$ nm half-width is obtained (Fig. S\ref{fig:S1}(c)) by bringing the
sample in the focal plane of the  
microscope while imaging the laser beam on a cooled CCD camera.
The PL is collected by exciting at $10~\mu$W of
optical power, and is filtered from the scattered laser light by a
high-energy-pass filter. The resulting PL image is shown in
Fig. S\ref{fig:S1}(d), clearly demonstrating the propagating character of
the emitting chiralitons. The logarithmic cross-cuts (red curves in (c)
and (d)) reveal a propagation length of several microns.

\begin{figure}
\includegraphics[width=0.8\columnwidth]{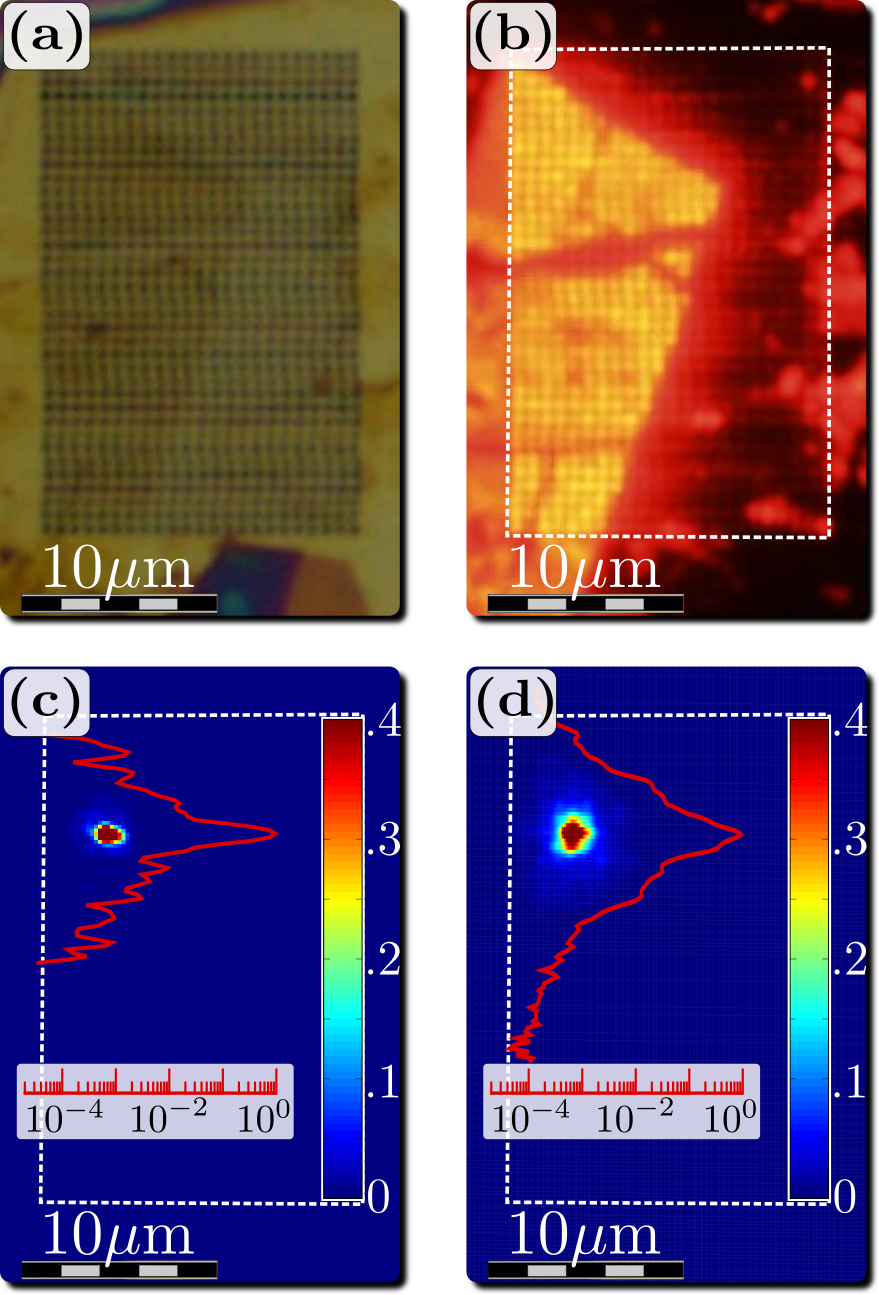}
\caption{(a) White light image of a WS$_2$ flake covering the
  plasmonic hole array, and (b) its wide field PL. (c) Normalized
  image of the diffraction-limited laser spot on the structure
  (indicated by the white dashed rectangle in (c) and (d)). The
  logarithmic scale cross-cuts (red curves in (c) and (d)) are taken
  along the vertical axis, through the intensity maximum. (d)
  Normalized chiraliton PL image.  \label{fig:S2}} 
\end{figure}

We note that the PL of the WS$_2$ monolayer also extends further away
from the flake above the OSO array (Fig. S\ref{fig:S2}(b)). We extract from this result the
$1/e$ decay length of the plasmon to be $\sim 3.4 ~\mu$m. This
value nicely agrees with that obtained in section A from the linear dispersion
analysis.

\section{C: Resonant Second Harmonic generation on a WS$_2$ monolayer}

\begin{figure}
\includegraphics[width=\columnwidth]{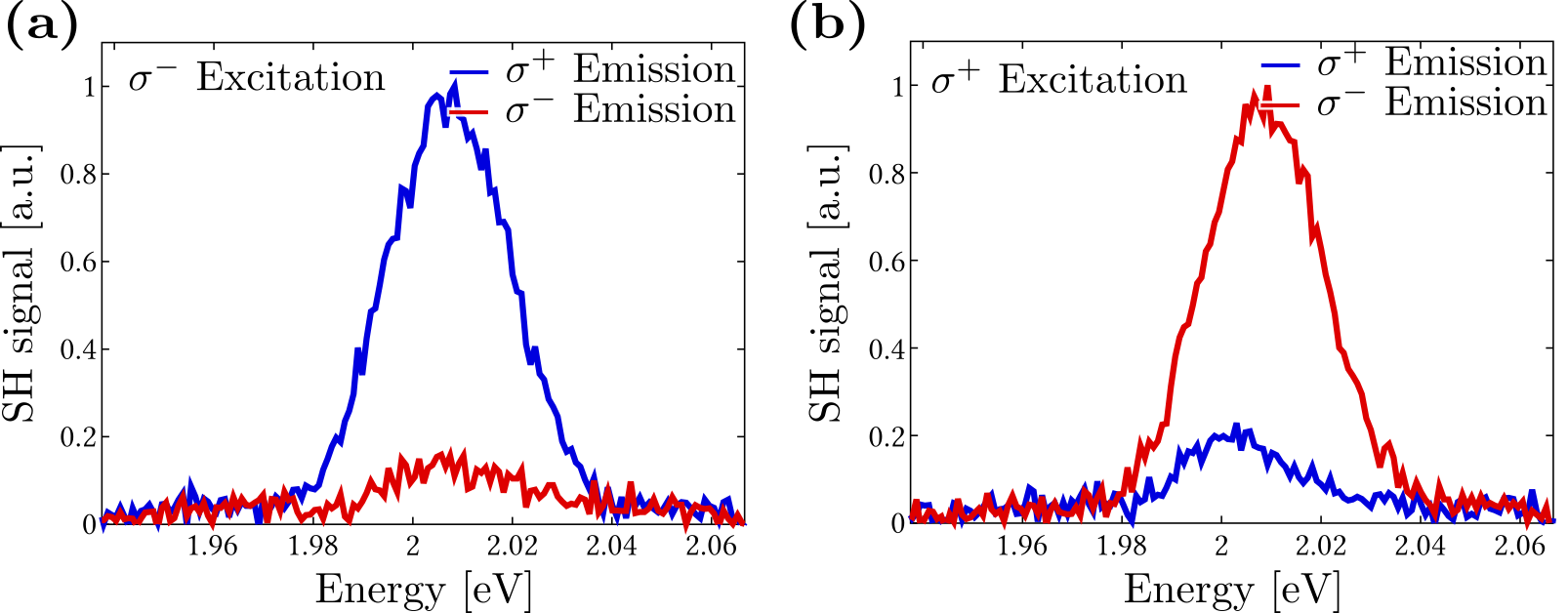}
\caption{Resonant SH spectrum  for left and right circular analysis,
  for (a) left and (b) right circular excitation at $1.01$eV. \label{fig:S3}} 
\end{figure}

As discussed in the main text, TMD monolayers have recently been
shown to give a high valley 
contrast in the generation of a second harmonic (SH) signal resonant
with their $A$-excitons. We obain a similar result when measuring a part
of the WS$_2$ monolayer sitting above   
the bare metallic surface, i.e. aside from the plasmonic hole array.
In Fig. S\ref{fig:S3} we show the SH signal obtained in left and right
circular polarization for an incident femto-second pump beam (120 fs
pulse duration, 1 kHz repetition rate at 1.01 eV) in (a) left and (b)
right circular polarization. 
This result confirms that the SH signal polarization is a good
observable of the valley degree of freedom of the WS$_2$ monolayer,
with a contrast reaching ca. $80\%$. 
In Fig.~3 (c) and (d) in the main text, we show how this valley contrast is
imprinted on the chiralitonic states. 

\section{D: Resonant SH generation in the strong coupling regime}

The resonant SH signal writes as \cite{Heinz1982,ChervyNano2016}:
\begin{equation}
I(2\omega)\propto (\rho_{\omega}I_{\omega})^2\cdot |\chi^{(2)}(2\omega) |^2 \cdot \rho_{2\omega}
\end{equation} 
where $I_{\omega}$ is the pump intensity, $\chi^{(2)}(2\omega)$ the second order susceptibility, $\rho_{\omega}$ the optical mode density of the resonator related to the fraction of the pump intensity that reaches WS$_2$ and $\rho_{2\omega}$ the optical mode density of the resonator that determines the fraction of SH intensity decoupled into the far field. While $\rho_{\omega}$ can safely be assumed to be non-dispersive at $\hbar\omega = 1$ eV, the dispersive nature of the resonator leads to $\rho_{2\omega}$ strongly dependent on the in-plane wave vector $k_x$. The optical mode density being proportional to the absorption, $\rho_{2\omega}(k_x)$ is given by the angular absorption spectrum crosscut at $2\hbar\omega = 2$ eV, displayed in the lower panels of Fig. 3 (e) and (f) in the main text.

Under the same approximations of \cite{LinPRA1993}, the resonant second order susceptibility can be written as
\begin{equation}
\chi^{(2)}(2\omega)=\alpha^{(1)}(2\omega)\sum_{n}\frac{K_{eng}}{\omega_{ng}-\omega}
\end{equation}
where $\sum_n$ sums over virtual electronic transitions, and $K_{eng}=\langle e|{\bf p}|g\rangle\otimes \langle e|{\bf p}|n\rangle\otimes\langle n|{\bf p}|g\rangle$ is a third-rank tensor built on the electronic dipole moments ${\bf p}$ taken between the $e,n,g$ states. The prefactor $\alpha^{(1)}(2\omega)$ is the linear polarizability of the system at frequency $2\omega$, yielding resonantly enhanced SH signal at every allowed $|g\rangle \rightarrow |e\rangle$ electronic transitions of the system. 

With two populations of uncoupled and strongly coupled WS$_2$ excitons, the SH signal is therefore expected to be resonantly enhanced when the SH frequency matches the transition frequency of either uncoupled or strongly coupled excitons. When the excited state is an uncoupled exciton associated with a transition energy fixed at frequency $2\hbar\omega=2$ eV for all angles, the tensor $K_{eng}$ is non-dispersive and the SH signal is simply determined and angularly distributed from $\rho_{2\omega}(k_x)$. 

When the excited state is a strongly coupled exciton, the resonant second order susceptibility becomes dispersive with $\chi^{(2)}(2\omega,k_x)$. This is due to the fact that the tensor $K_{eng}$ incorporates the excitonic Hopfield coefficient of the polaritonic state involved in the electronic transition $|g\rangle \rightarrow |e\rangle$ when the excited state is a polaritonic state. In our experimental conditions with a pump frequency at $1$ eV, this excited state is the upper polaritonic state with $|e\rangle \equiv |P^+_{K(K'),\sigma^\pm},\mp k_{\rm SP}\rangle$ and therefore $K_{eng}\propto [\beta_{K(K')}^+ (k_x)]^2$. This dispersive excitonic Hopfield coefficient is evaluated by the procedure described in details above, Sec. A. The profile of the SH signal then follows the product between the optical mode density $\rho_{2\omega}(k_x)$ and $|\chi^{(2)}(2\omega,k_x)|^2\propto[\beta_{K(K')}^+ (k_x)]^4$.

These two contributions are perfectly resolved in the SH data displayed in Fig. 3 (e) and (f) in the main text. The angular distribution of the main SH signal clearly departs from $\rho_{2\omega}(k_x)$, revealing the dispersive influence of $\beta_{K(K')}^+ (k_x)$. This is perfectly seen on the crosscuts displayed in the lower panels of Figs.~3 (e) and (f) in the main text. This feature is thus an indisputable proof of the existence of chiralitonic states, i.e. of the strongly coupled nature of our system. 

A residual SH signal is also measured which corresponds to the contribution of uncoupled excitons. This residual signal is measured in particular within the anticrossing region, as expected from the angular profile of $\rho_{2\omega}(k_x)$ shown in the lower panels of Figs.~3 (e) and (f) in the main text.

Finally, the angular features of the SH signal exchanged when the spin of the pump laser is flipped from $\sigma^+$ to $\sigma^-$ reveal how valley contrasts have been transferred to the polariton states. These features therefore demonstrate the chiral nature of the strong coupling regime, i.e. the existence of genuine \textit{chiralitons}.

\section{E: PL lifetime measurement on the strongly coupled system}

\begin{figure}
\includegraphics[width=\columnwidth]{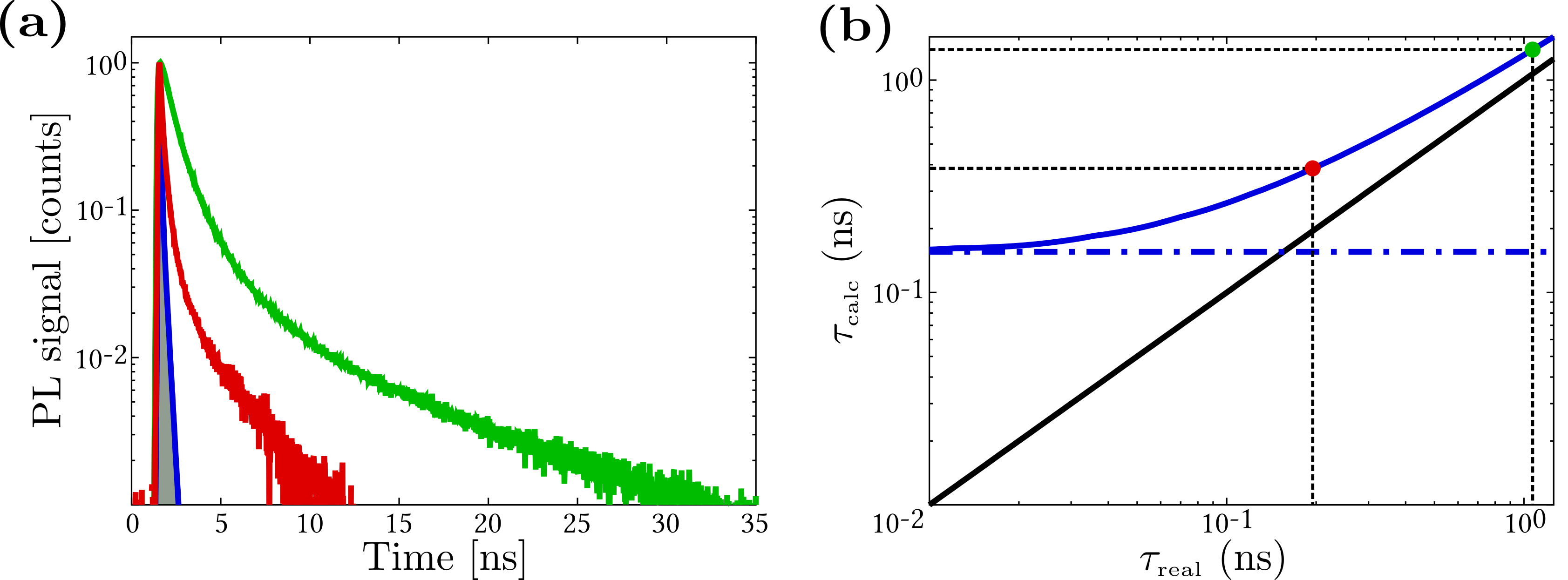}
\caption{(a) TCSPC histogram showing the PL decay dynamic of the
  strongly coupled WS$_2$ monolayer (red curve), as compared to that of a
  bare WS$_2$ monolayer on a dielectric substrate (green curve). The
  IRF of our measurement apparatus is shown in blue. (b) Calibration
  (blue curve) used to retrieve 
  $\tau_{\textrm{\tiny real}}$ from the measurement of
  $\tau_{\textrm{\tiny calc}}$, obtained by convoluting an exponential
  decay of time constant  $\tau_{\textrm{\tiny real}}$ with the
  measured IRF. The calculated IRF time constant
  $\tau_{\textrm{\tiny calc}}^{\textrm{\tiny IRF}} = 157$ ps is shown
  by the blue dashed line. $\tau_{\textrm{\tiny calc}}^{\textrm{\tiny ref}}$ and  
  $\tau_{\textrm{\tiny calc}}^{\textrm{\tiny sample}}$ are represented
  as green and red spots respectively.
  \label{fig:S4}} 
\end{figure}

The PL lifetime of the strongly coupled system is measured by
time-correlated single photon counting (TCSPC) under pico-second pulsed
excitation (instrument response time $120$ ps, 20 MHz repetition rate at 1.94 eV).
The arrival time histogram of PL photons, when measuring a part of the
WS$_2$ monolayer located above the plasmonic hole array, gives the decay dynamic shown
in Fig. S\ref{fig:S4}(a). On this figure we also display the PL decay of a
reference WS$_2$ monolayer exfoliated on a dielectric substrate
(polydimethylsiloxane), as well as the instrument response function
(IRF) measured by recording the excitation pulse photons scattered by a gold
film. Following the procedure detailed in \cite{Berciaud2015}, we
define the {\it calculated} decay times $\tau_{\textrm{\tiny calc}}$ as the
area under the decay curves (corrected for their backgrounds) divided
by their peak values. This yields a calculated IRF time constant
$\tau_{\textrm{\tiny calc}}^{\textrm{\tiny IRF}} = 157$ ps, and 
calculated PL decay constants $\tau_{\textrm{\tiny
calc}}^{\textrm{\tiny ref}} = 1.39$ ns and  $\tau_{\textrm{\tiny
calc}}^{\textrm{\tiny sample}} = 384$ ps for the reference bare flake
and the strongly coupled sample respectively. The {\it real}
decay time constants  $\tau_{\textrm{\tiny
real}}$ corresponding to the calculated ones can then be
estimated by convoluting different monoexponential decays with the
measured IRF, computing the corresponding $\tau_{\textrm{\tiny
calc}}$ and interpolating this calibration curve
(Fig. S\ref{fig:S3}(b)) for the values of $\tau_{\textrm{\tiny
calc}}^{\textrm{\tiny ref}}$ and  $\tau_{\textrm{\tiny
calc}}^{\textrm{\tiny sample}}$. This results in $\tau_{\textrm{\tiny
real}}^{\textrm{\tiny ref}} = 1.06$ ns and  $\tau_{\textrm{\tiny
real}}^{\textrm{\tiny sample}} = 192$ ps.

While the long life-time (ns) of the bare WS$_2$ exciton has been attributed to the trapping of the exciton outside the light-cone at RT through phonon scattering \cite{PollmannNatMat2015,RobertPRB2016}, Fig.~S\ref{fig:S4} simply shows that the exciton life time is reduced by the presence of the metal or the OSO resonator. Clearly, the competition induced under strong coupling conditions between the intra and inter valley relaxation rates and the Rabi oscillations must act under shorter time scales that are not resolved here.

\section{F: Optical setup}

\begin{figure}
\includegraphics[width=\columnwidth]{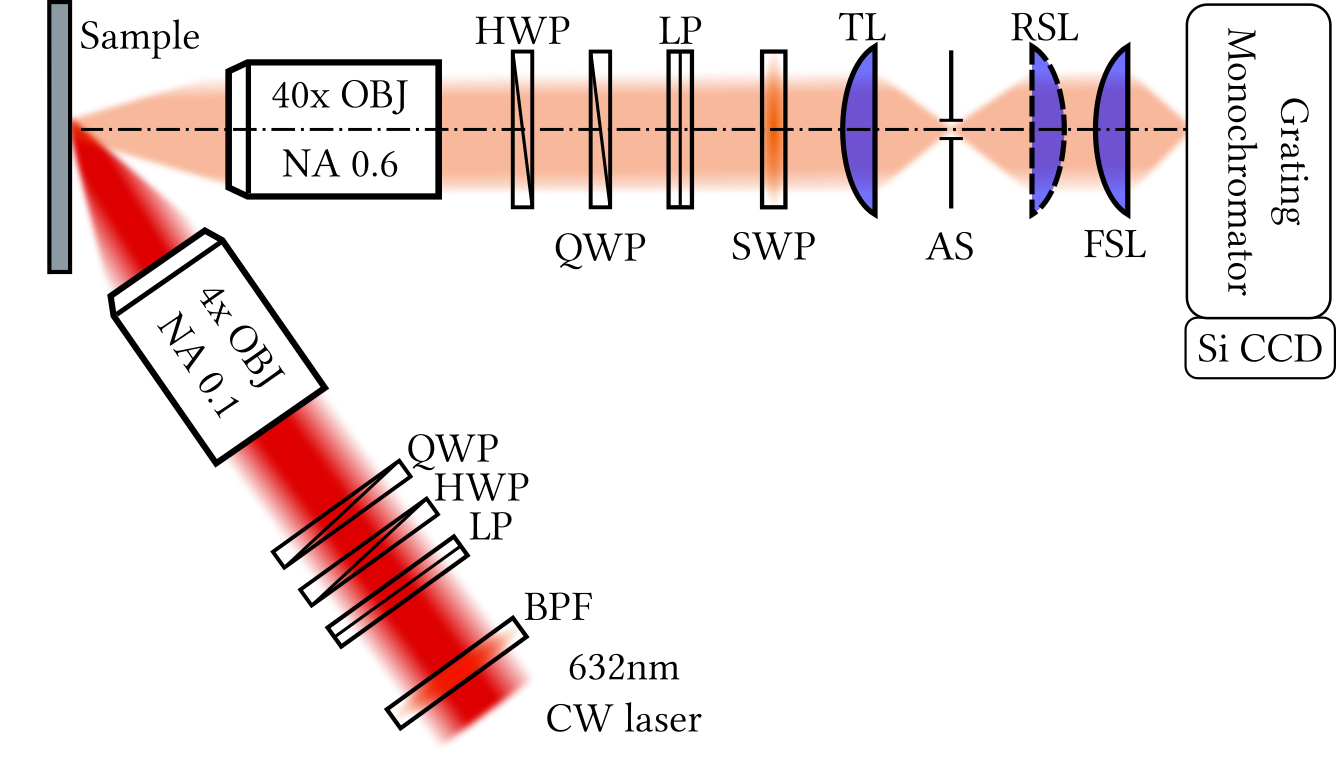}
\caption{Optical setup used for the angle-resolved polarimetric
  measurements. See the corresponding paragraph for details.
  \label{fig:S5}} 
\end{figure}

The optical setup used for PL polarimetry
experiments is shown in Fig. S\ref{fig:S5}. The WS$_2$
monolayer is excited by a continous-wave HeNe laser at $1.96$ eV (632 nm),
slightly below the direct band-gap of the atomic crystal, in order to
reduce phonon-induced inter-valley scattering effects at room
temperature. The pumping laser beam is filtered by a bandpass filter
(BPF) and its polarization state is controlled by a set of
polarization optics: a linear polarizer (LP), a half-wave plate
(HWP) and a quarter-wave plate (QWP). The beam is focused onto
the sample surface at oblique incidence angle by a microscope
objective, to a typical spot size of $100~\mu\textrm{m}^2$. This corresponds to
a typical flux of $10$ W$\cdot$cm$^{-2}$. In such conditions of
irradiation, the PL only comes from the $A-$exciton. The emitted  
PL signal is collected by a high 
numerical aperture objective, and its polarization state is analyzed
by another set of broadband polarization optics (HWP, QWP, LP). A
short-wavelength-pass (SWP) tunable filter is placed on the
optical path to stop the laser light scattered. Adjustable
slits (AS) placed at the image plane of the 
tube lens (TL) allow to spatially select the PL signal coming only
from a desired area of the sample, whose Fourier-space (or real space) spectral
content can be imaged onto the entrance slits of the spectrometer by a
Fourier-space lens (FSL), or adding a real-space lens (RSL). The resulting image is recorded by a cooled CCD Si camera.

\section{G: Valley contrast measurements on a bare WS$_2$ monolayer}

\begin{figure}
\includegraphics[width=\columnwidth]{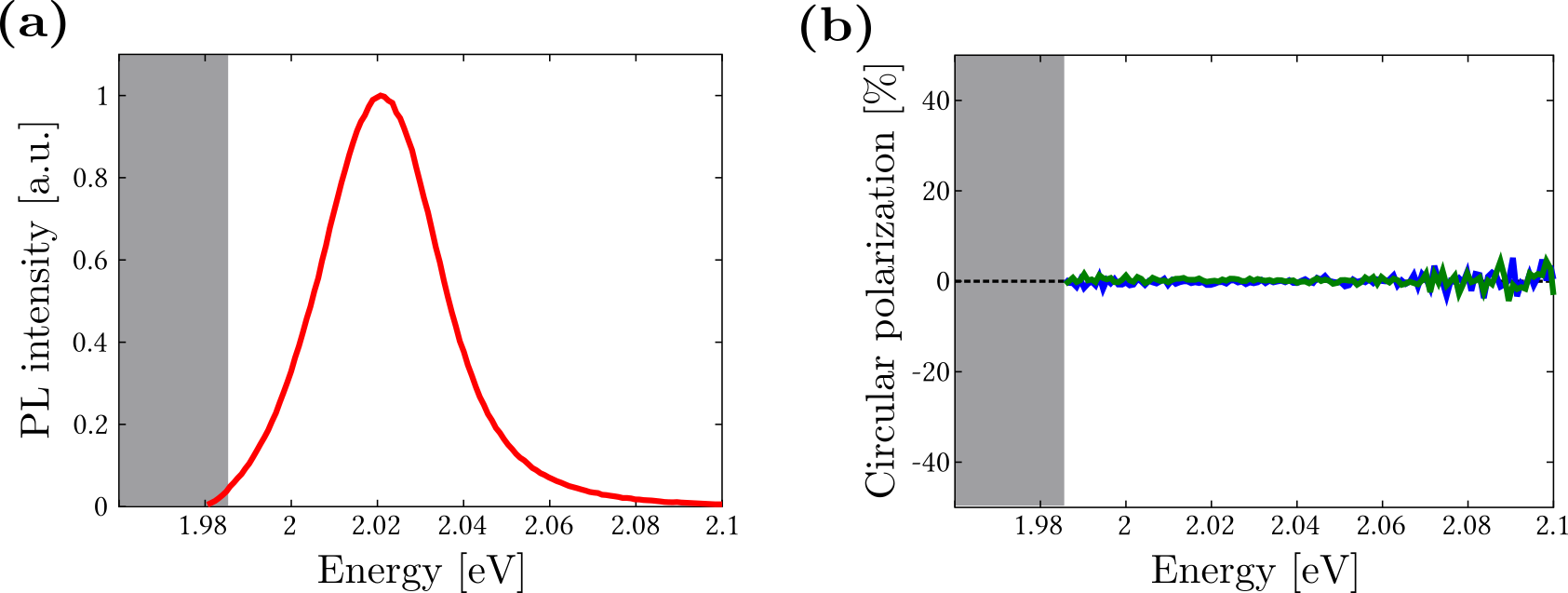}
\caption{(a) Emission spectrum $I_{\sigma^-}(\sigma^-)$ obtained by exciting the bare WS$_2$
  monolayer at $1.96$ eV with $\sigma^-$ polarized light, and
  analysing the PL in $\sigma^-$ polarization. The gray area in (a)
  and (b) corresponds to the spectral region 
  cut by the PL emission filter. (b) The
  valley contrast  $\rho^{+(-)}$ defined in (\ref{eq:CPL}) is displayed in
  blue (green).
  \label{fig:S6}} 
\end{figure}

The valley contrast $\rho^\pm$ of a  bare WS$_2$ monolayer exfoliated on a
dielectric substrate (polydimethylsiloxane) is computed from the
measured room temperature PL spectra obtained for left and right circular excitations,
analysed in the circular basis by a combination of a
quarter-wave plate and a Wollaston prism:
\begin{equation}
  \rho^\pm = \frac{I_{\sigma^\pm}(\sigma^+) - I_{\sigma^\pm}(\sigma^-)}{I_{\sigma^\pm}(\sigma^+) + I_{\sigma^\pm}(\sigma^-)},
 \label{eq:CPL}
\end{equation}
where $I_j(l)$ is the measured PL spectrum for a
$j=(\sigma^+,\sigma^-)$ polarized excitation and a
$l=(\sigma^+,\sigma^-)$ polarized analysis.
A typical emission spectrum ($I_{\sigma^-}(\sigma^-)$) is shown in
Fig. S\ref{fig:S6}(a) and the valley contrasts  $\rho^\pm$ are displayed
in Fig. S\ref{fig:S6}(b). As discussed in the main text, this emission
spectrum consists of a phonon-induced up-converted PL \cite{Xiaodong2016}.
Clearly, there is no difference in the
$I_j(l)$ spectra, hence no valley polarization at room temperature on
the bare  WS$_2$ monolayer. These results are in striking contrast to those
reported in the main text for the strongly coupled system, under
similar excitation conditions. Note also
that the absence of valley contrast on our bare WS$_2$ monolayer
differs from the results of \cite{PoWenChiu2016} reported however on WS$_2$
grown by chemical vapor deposition.

\section{H: Angle-resolved Stokes vector polarimetry}

The optical setup shown in Fig. S\ref{fig:S5} is used to measure the
angle-resolve PL spectra for different combinations of excitation and
detection polarizations. Such measurements allow us to retrieve the
coefficients of the Mueller matrix $\mathcal{M}$ of the system, characterizing how the
polarization state of the excitation beam affects the polarization
state of the chiralitons PL. As discussed in the main text, the
spin-momentum locking mechanism of our chiralitonic system
relates such PL polarization states to
specific chiraliton dynamics. An incident excitation in a given polarization state
is defined by a Stokes vector $\bf{S}^{\textrm{\tiny in}}$, on which
the matrix $\mathcal{M}$ acts 
to yield an output PL Stokes vector $\bf{S}^{\textrm{\tiny out}}$:
\begin{equation}
 \bf{S}^{\textrm{\tiny out}} = \left(
   \begin{array}{c}
     I\\
     I_{V} - I_{H}\\
     I_{45} - I_{-45}\\
     I_{\sigma^+} - I_{\sigma^-}
   \end{array}
 \right)_{\textrm{\tiny out}} = \mathcal{M}\left(
   \begin{array}{c}
     I_0\\
     I_{V} - I_{H}\\
     I_{45} - I_{-45}\\
     I_{\sigma^+} - I_{\sigma^-}
   \end{array}
 \right)_{\textrm{\tiny in}}, \label{eq:Mueller}
\end{equation}
where $I_{(0)}$ is the emitted (incident) intensity, $I_{V} - I_{H}$
is the relative intensity in vertical and horizontal polarizations, $I_{45} - I_{-45}$
is the relative intensity in $+45^o$ and  $-45^o$ polarizations and 
$I_{\sigma^+} - I_{\sigma^-}$ is the relative intensity in $\sigma^+$
and  $\sigma^-$ polarizations.
We recall that for our specific alignment of the OSO resonator with
respect to the slits of the spectrometer, the angle-resolved PL
spectra in $V$ and $H$ polarizations correspond to
transverse-magnetic (TM) and transverse-electric (TE) dispersions
respectively (see Fig.~2 (b) in the main text). 
Intervalley chiraliton coherences, revealed by a non-zero
degree of linear polarization in the PL upon the same linear
excitation, are then measured by the $S_1 = I_{V} - I_{H}$ coefficient of the PL
output Stokes vector. This coefficient is obtained
by analysing the PL in the linear basis, giving an angle-resolved PL
intensity $\left(S_0^{\textrm{\tiny out}} +(-) S_1^{\textrm{\tiny out}}\right)/2$,
for TM (TE) analysis. In order to obtain the polarization
characteristics of the chiralitons, we measure the four possible
combinations of excitation and detection polarization in the linear
basis:
\begin{eqnarray}
  I_{\textrm{\tiny TM/TM}} &=& \left(m_{00} + m_{01} + m_{10} + m_{11}\right)/2\\
  I_{\textrm{\tiny TM/TE}} &=& \left(m_{00} + m_{01} - m_{10} - m_{11}\right)/2\\
  I_{\textrm{\tiny TE/TM}} &=& \left(m_{00} - m_{01} + m_{10} - m_{11}\right)/2\\
  I_{\textrm{\tiny TE/TE}} &=& \left(m_{00} - m_{01} - m_{10} + m_{11}\right)/2,  
\end{eqnarray}
where $I_{p/a}$ is the angle-frequency resolved intensity measured for
a pump polarization $p=(\textrm{TE,TM})$ and analysed in
$a=(\textrm{TE,TM})$ polarization, and $m_{i,j}$ are the coefficients
of the 4x4 matrix $\mathcal{M}$. By solving this linear system of
equations, we obtain the first quadrant of the Mueller matrix: $m_{00},
m_{01}, m_{10}$ and $m_{11}$.
The $S_1^{\textrm{\tiny out}}|_{\textrm{\tiny TM}}$ coefficient of the output Stokes vector
for a TM excitation is then directly given by $m_{10}+m_{11}$ as
can be seen from (\ref{eq:Mueller}) by setting $I_V=1, I_H=0$ and all
the other input Stokes coefficients to zeros. This quantity,
normalized to $S_0^{\textrm{\tiny out}}$, is
displayed in the $k_x-$energy plane in Fig.~4 (c) in the main text. Similarly, the
$S_1^{\textrm{\tiny out}}|_{\textrm{\tiny TE}}$ coefficient is given by  
$m_{10}-m_{11}$, which is the quantity displayed in Fig.~4 (d) in the main text.

As the dispersion of the OSO resonator is different for TE and TM
polarizations, the pixel-to-pixel operations performed to obtain
$S_1^{\textrm{\tiny out}}$ do not directly yield the chiraliton
inter-valley contrast. In particular, the observation of negative
value regions in $S_1^{\textrm{\tiny out}}|_{\textrm{\tiny TM}}$ only reveals that the
part of the chiraliton population that lost inter-valley
coherence is dominating the total PL in the region of the dispersion where
the TE mode dominates over the TM mode (compare Fig.~4(c) and (e)). It
{\it does not} correspond to genuine anti-correlation of the chiraliton PL polarization
with respect to the pump polarization. To correct for such dispersive
effects and obtain the degree of chiraliton intervalley coherence, the
appropriate quantity is $(S_1^{\textrm{\tiny out}}|_{\textrm{\tiny TM}}-S_1^{\textrm{\tiny out}}|_{\textrm{\tiny TE}})/(2S_0^{\textrm{\tiny out}}) = m_{11}$,
resolved in the $k_x-$energy plane in
Fig.~4 (f) in the main text. This quantity can also be refered to as
a chiraliton linear depolarization factor.
For these polarimetry measurements, the base-line noise was determined by measuring the Mueller matrix associated with an empty setup which are expected to be proportional to the identity matrix. With polarizer extinction coefficients smaller than $0.1\%$, white light (small) intensity fluctuations, and positioning errors of the polarization optics, we reach standard deviations from the identity matrix of the order of $0.4\%$. This corresponds to a base-line noise valid for all the polarimetry measurements presented in the main text. The noise level seen in Fig. 4 in the main text is thus mostly due to fluctuations in the WS$_2$ PL intensity.

\end{document}